# Inductive Logic
# From Data Analysis to Experimental Design

Kevin H. Knuth

*Center for Advanced Brain Imaging*
*Nathan Kline Institute, Orangeburg NY 10962*

**Abstract.** In celebration of the work of Richard Threlkeld Cox, we explore inductive logic and its role in science touching on both experimental design and analysis of experimental results. In this exploration we demonstrate that the duality between the logic of assertions and the logic of questions has important consequences. We discuss the conjecture that the relevance or bearing, *b*, of a question on an issue can be expressed in terms of the probabilities, *p*, of the assertions that answer the question via the entropy.

In its application to the scientific method, the logic of questions, inductive inquiry, can be applied to design an experiment that most effectively addresses a scientific issue. This is performed by maximizing the relevance of the experimental question to the scientific issue to be resolved. It is shown that these results are related to the mutual information between the experiment and the scientific issue, and that experimental design is akin to designing a communication channel that most efficiently communicates information relevant to the scientific issue to the experimenter. Application of the logic of assertions, inductive inference (Bayesian inference) completes the experimental process by allowing the researcher to make inferences based on the information obtained from the experiment.

## THE LOGIC OF INFERENCE AND INQUIRY

These workshops have spanned over two decades of research during which the power of Bayesian (or inductive) inference has been demonstrated time and time again. Slowly, but surely, these techniques have become more accepted in mainstream science with applications in virtually every field. Even as I write, the Office Assistant on this word processor, which uses a Bayesian network to infer my intentions from my actions is offering a suggestion to help me with the formatting of this document. It is performing the equivalent of data analysis, which is arriving at the most probable conclusions given one's prior knowledge and newly acquired data.

While data analysis is an extremely important part of scientific investigation, its counterpart, experimental design is equally important. Intuitively, the problem of experimental design, which consists of choosing an experimental question most relevant to the scientific issue to be resolved, is related to data analysis. However, there does not yet exist a complete theory of the logic of inference and inquiry. The goal of this paper is to introduce the reader to the overarching framework of inductive logic, to describe what is known regarding the relationships between inference, inquiry, probability theory and information theory, and to highlight what is not known.

# Deductive and Inductive Inference

As deductive inference refers to implication among logical assertions in situations of complete certainty, we begin with Boolean logic. An assertion $a$ implies an assertion $b$, written $a \rightarrow b$, if $a \wedge b = a$ and $a \vee b = b$, where $\wedge$ is the logical *and* operation such that $a \wedge b$ is an assertion that tells what $a$ and $b$ tell jointly, and $\vee$ is the logical *or* operation such that $a \vee b$ is an assertion that tells what $a$ and $b$ tell in common. As an example consider the two assertions $a = $"*It is a Kangaroo!*" and $b = $"*It is an Animal!*".[1] The assertion $a$ implies the assertion $b$ as jointly the two assertions say *"It is a Kangaroo!"*. In addition, the common assertion $a \vee b$ says *"It is an Animal!"*. Table 1 (below) lists the Boolean identities for assertions.

Richard T. Cox's major contribution [1,2] to inductive inference arises from generalizing Boolean implication to implication of varying degree, where the real number representing the degree to which the implicate $b$ is implied by the implicant $a$ is written as $(a \rightarrow b)$. The inferential utility of this formalism is readily apparent when the implicant is an assertion representing a premise and the implicate is an assertion representing a hypothesis.

From the associativity of the conjunction of assertions, $(a \rightarrow (b \wedge c) \wedge d) = (a \rightarrow b \wedge (c \wedge d))$, Cox derived a functional equation, which has as a particular solution

$$(a \rightarrow b \wedge c) = (a \rightarrow b)(a \wedge b \rightarrow c). \qquad (1)$$

In addition, if you know something about an assertion, you also know something about its contradictory. In other words, the degree to which a premise implies an assertion $b$ determines the degree to which the premise implies its contradictory ~$b$. This logical principle can be applied twice to obtain a functional equation, which has as a particular solution

$$(a \rightarrow b) + (a \rightarrow \sim b) = 1. \qquad (2)$$

In general the first functional equation puts some constraints on the second, which results in a general solution

$$(a \rightarrow b \wedge c)^r = (a \rightarrow b)^r (a \wedge b \rightarrow c)^r \qquad (3)$$

$$(a \rightarrow b)^r + (a \rightarrow \sim b)^r = C, \qquad (4)$$

where $r$ and $C$ are arbitrary constants. Setting $r = C = 1$ one obtains the particular solutions above.

Cox demonstrated that this measure of relative degree of implication among assertions is the unique logically consistent measure. We do well to define probability as this relative degree of implication among assertions. In fact, a simple change of notation $p(b|a) \equiv (a \rightarrow b)$ reveals that the equations (1) and (2) above

---

[1] Here we adopt the notation used by Cox where an assertion is denoted by a lowercase Roman character, and a question is denoted by an uppercase Roman character. In addition, we adopt the notation used by Fry where assertions are stated with exclamation marks and questions with question marks.

$$p(b \wedge c \mid a) = p(b \mid a) \, p(c \mid a \wedge b) \qquad (5)$$
$$p(b \mid a) + p(\sim b \mid a) = 1 \qquad (6)$$

are the product and sum rules, respectively, of probability theory.

Utilizing the commutativity of the conjunction of two assertions $b \wedge c \equiv c \wedge b$, equation (5) can be applied to obtain

$$p(b \wedge c \mid a) = p(b \mid a) \, p(c \mid a \wedge b) \qquad (7)$$
$$p(c \wedge b \mid a) = p(c \mid a) \, p(b \mid a \wedge c). \qquad (8)$$

Equating the right-hand sides of (7) and (8), we obtains Bayes' Theorem

$$p(b \mid a \wedge c) = p(b \mid a) \frac{p(c \mid a \wedge b)}{p(c \mid a)}, \qquad (9)$$

which allows one to evaluate the probability of a hypothesis given one's prior knowledge and newly acquired data. The foundation of data analysis rests on this theorem.

Two important points should be noted. First, this formalism allows one to perform inductive inference over a broad range of applications. Given a set of assertions this calculus allows one to determine the relative degree to which any assertion implies any other. This is far beyond the scope supported by frequentist statistics. Second, there cannot be implication without an implicant. In short, probabilities are always conditional on some state of prior knowledge.

## Deductive and Inductive Inquiry

While it is possible to examine the logical relationships among what is known, it is equally possible to examine the logic of what is unknown. Cox's second major contribution [3] was to lay the foundations for the logic of questions. He defined a question as the set of assertions that answer the question. For example, the question $K = $ "*In what state does my kangaroo live?*" can be expressed in terms of assertions by

$$K = \begin{cases} k1 = \text{"}Tasmania!\text{"}, & k2 = \text{"}New\ South\ Wales!\text{"}, \\ k3 = \text{"}Queensland!\text{"}, & k4 = \text{"}Western\ Australia!\text{"}, \\ k5 = \text{"}Victoria!\text{"}, & k6 = \text{"}South\ Australia!\text{"}, \\ & k7 = \text{"}Northern\ Territory!\text{"} \end{cases}. \qquad (10)$$

This defining set of assertions can be extended without changing the question by including assertions like $k1 \wedge k3$, as this conjunction implies $k1$ and $k3$, which are already in the set. A system of assertions is a set, which includes every assertion implying any assertion in the set. The irreducible set is a subset of the system such that no assertion in the irreducible set implies any other in that set, except itself.

The conjunction of two questions is called the joint question. It asks what the two questions ask jointly. In terms of assertions, the joint question can be written as

$$A \wedge B = \begin{Bmatrix} a1 \wedge b1, & a1 \wedge b2, & \cdots & a1 \wedge bm, \\ a2 \wedge b1, & \ddots & & \vdots \\ \vdots & & & \\ an \wedge b1, & \cdots & & an \wedge bm \end{Bmatrix}, \quad (11)$$

which is not a matrix, but a set of all possible pairs of conjunctions of the assertions defining the questions $A$ and $B$. Similarly, the disjunction of two questions, called the common question, is defined as the question that the two questions ask in common. In terms of assertions it can be answered by the union of the sets of assertions answering each question

$$A \vee B = \{a1, a2, \cdots, an, b1, b2, \cdots, bm\}. \quad (12)$$

With these definitions, one can derive the Boolean identities for questions shown in Table 1. Note that they are symmetric with the relations for assertions under interchange of disjunction $\vee$ and conjunction $\wedge$.

| TABLE 1. Boolean Identities | | |
|---|---|---|
| | **Assertions** | |
| A1 | $a \wedge a = a$ | $a \vee a = a$ |
| A2 | $a \wedge b = b \wedge a$ | $a \vee b = b \vee a$ |
| A3 | $(a \wedge b) \wedge c = a \wedge b \wedge c$ | $(a \vee b) \vee c = a \vee b \vee c$ |
| A4 | $(a \wedge b) \vee c = (a \vee c) \wedge (b \vee c)$ | $(a \vee b) \wedge c = (a \wedge c) \vee (b \wedge c)$ |
| A5 | $(a \wedge b) \vee b = b$ | $(a \vee b) \wedge b = b$ |
| A6 | $\sim \sim a = a$ | |
| A7 | $(a \wedge \sim a) \wedge b = a \wedge \sim a$ | $(a \vee \sim a) \vee b = a \vee \sim a$ |
| A8 | $(a \wedge \sim a) \vee b = b$ | $(a \vee \sim a) \wedge b = b$ |
| A9 | $\sim (a \wedge b) = \sim a \vee \sim b$ | $\sim (a \vee b) = \sim a \wedge \sim b$ |
| | **Questions** | |
| Q1 | $A \vee A = A$ | $A \wedge A = A$ |
| Q2 | $A \vee B = B \vee A$ | $A \wedge B = B \wedge A$ |
| Q3 | $(A \vee B) \vee C = A \vee B \vee C$ | $(A \wedge B) \wedge C = A \wedge B \wedge C$ |
| Q4 | $(A \vee B) \wedge C = (A \wedge C) \vee (B \wedge C)$ | $(A \wedge B) \vee C = (A \vee C) \wedge (B \vee C)$ |
| Q5 | $(A \vee B) \wedge B = B$ | $(A \wedge B) \vee B = B$ |
| Q6 | $\sim \sim A = A$ | |
| Q7 | $(A \vee \sim A) \vee B = A \vee \sim A$ | $(A \wedge \sim A) \wedge B = A \wedge \sim A$ |
| Q8 | $(A \vee \sim A) \wedge B = B$ | $(A \wedge \sim A) \vee B = B$ |
| Q9 | $\sim (A \vee B) = \sim A \wedge \sim B$ | $\sim (A \wedge B) = \sim A \vee \sim B$ |

Analogous to implication among assertions, one can define an ordering relation on questions, which we shall call inclusion[2], such that a question $A$ includes question $B$, written $A \rightarrow B$, if $A \wedge B = A$ and $A \vee B = B$. This can be more easily visualized by considering $A =$ "*What kind of animal is it ?*" and $B =$ "*Is it a Kangaroo or not ?*". Jointly, the questions ask $A \wedge B =$ "*What kind of animal is it ?*" and in common the questions ask $A \vee B =$ "*Is it a Kangaroo or not ?*". Thus question $A$ includes question $B$. This can also be verified by considering the questions in terms of the set of assertions that answer them.

---

[2] It was suggested by Anton Garrett at the MaxEnt 2000 workshop that the relation $A \rightarrow B$ be read as $A$ includes $B$.

Inclusion can be generalized from a binary relation to a degree of inclusion represented by a real number $(A \rightarrow B)$. This real number can be thought of as describing the bearing that question $A$ has on issue $B$ or the relevance[3] of question $A$ on issue $B$. Adopting a notational change, we denote this function as $b(A | B) \equiv (A \rightarrow B)$.[4] Note that the position of the questions relative to the solidus "|" is opposite of that for the definition of probability, $p(b | a) \equiv (a \rightarrow b)$.

Just as with assertions, one can derive the sum and product rules for relevance

$$b(A \vee B | C) = b(A | B \vee C) b(B | C) \qquad (13)$$
$$b(A | B) + b(\sim A | B) = 1. \qquad (14)$$

The commutativity of the disjunction operation can be used with the product rule to derive the equivalent of Bayes' Theorem for questions

$$b(B | A \vee C) = b(B | C) \frac{b(A | B \vee C)}{b(A | C)}. \qquad (15)$$

## RELATIONSHIPS

In this section we examine some relationships between the logic of assertions and the logic of questions. In doing so we shed new light on the relationship between Bayesian inference and information theory.

### Bayes Theorem for Assertions, Entropy and Information Theory

We begin with the product rule (as we did when deriving Bayes' Theorem):

$$p(b \wedge c | h) = p(b | h) \, p(c | b \wedge h) \qquad (16)$$
$$p(c \wedge b | h) = p(c | h) \, p(b | h \wedge c), \qquad (17)$$

where the assertion $h$ is a joint or compound assertion representing all that is known. Taking the logarithm of both sides of equation (16)

$$\log p(b \wedge c | h) \;=\; \log p(b | h) \;+\; \log p(c | b \wedge h), \qquad (18)$$

and taking the expected value over all possible assertions $b \wedge c$ we find

---

[3] The term 'bearing' was adopted by Robert Fry who saw its long-standing use in English law as being appropriately descriptive. However it was brought up at the meeting that this term may be difficult for non-English speakers who do not have an obvious equivalent in their language and thus may find this term obscure. The term 'relevance' was suggested as an alternative.

[4] Regardless of whether the term 'bearing' or 'relevance' is used, we adopt Robert Fry's notation for this function, which is based on the term 'bearing'. This notation is especially pleasing as the letter 'b' is an upside-down 'p' (for probability), which highlights the symmetry between the logic of assertions and the logic of questions.

$$\sum_{b,c} p(b \wedge c \mid h) \log p(b \wedge c \mid h)$$
$$= \sum_{b,c} p(b \wedge c \mid h) \log p(b \mid h) + \sum_{b,c} p(b \wedge c \mid h) \log p(c \mid b \wedge h) \quad (19)$$

By the sum rule of probability, the sum over $c$ in the first term on the right-hand side marginalizes to one leaving only the sum over $b$. One can easily see that each term is negative one times some entropy. More specifically, one can write (19) as

$$H(b,c) = H(b) + H(c \mid b), \quad (20)$$

where $H(b, c)$ is defined in information theory as the joint entropy of $b$ and $c$, $H(b)$ is the entropy of $b$, and $H(c \mid b)$ is the conditional entropy of $c$ given $b$. Application of the same procedure to (17), equating the right-hand sides and solving for $H(b \mid c)$ gives

$$H(b \mid c) = H(b) + H(c \mid b) - H(c), \quad (21)$$

which is the information-theoretic equivalent of Bayes' Theorem

$$p(b \mid c \wedge h) = p(b \mid h) \frac{p(c \mid b \wedge h)}{p(c \mid h)}. \quad (22)$$

With this in mind, the notation adopted by information theory is quite pleasing as one can easily visualize the correspondence between equations (21) and (22).

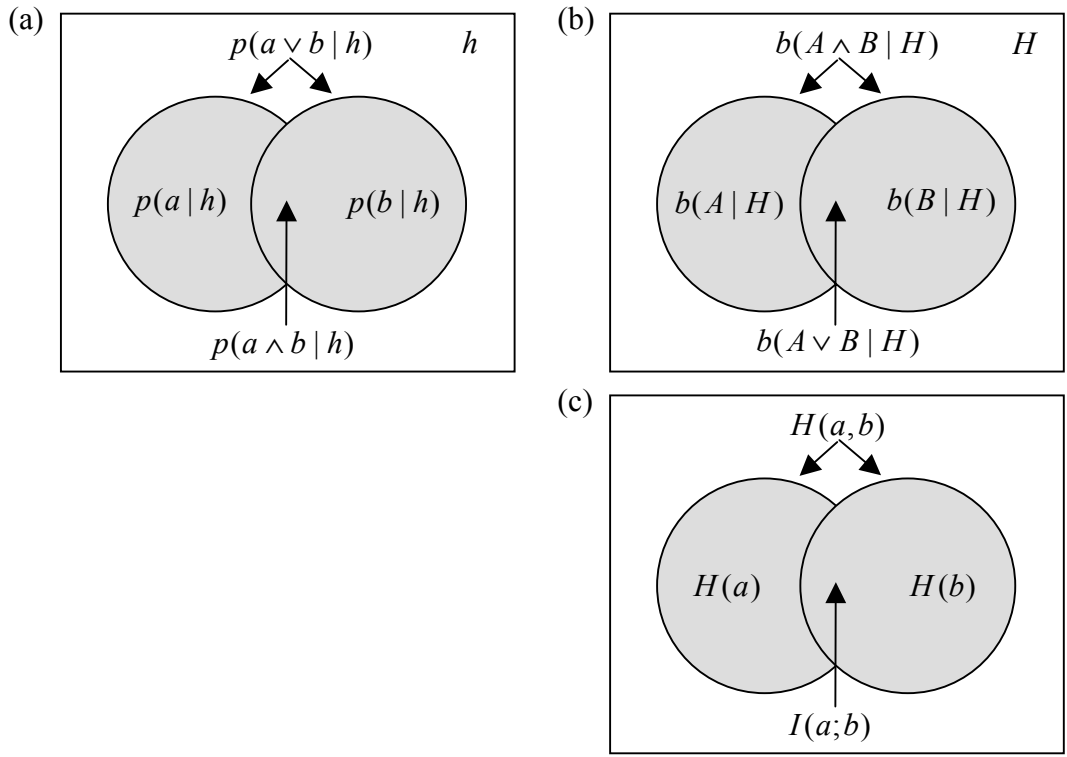

**FIGURE 1.** Venn diagrams demonstrating the symmetries between (a) the logic of assertions and (b) the logic of questions. (c) An *I*-diagram representing the analogous situation in information theory. Note that in the *I*-diagram the function $H(\cdot)$ is entropy.

## The Relevance - Entropy Conjecture

As questions can be defined in terms of assertions, one would expect that the relevance or bearing of one question on another could be expressed in terms of the assertions that answer those questions. This should depend on the probability of (or degree of implication among) those assertions that answer the questions. The symmetries between the Venn diagram for two questions (Figure 1b) and the *I*-diagram for entropy in information theory (Figure 1c) suggest strongly that entropy is the appropriate measure of relevance in terms of probability [4]. In addition, Cox [3] demonstrates that the properties of entropy seem to make it a convenient measure of relevance. However, as no proof yet exists, it is still only conjecture that the relevance or bearing of a question on an issue can be written as the entropy of the probabilities of the assertions that answer those questions.

It is interesting to rewrite equation (21) in terms of the relevance assuming the conjecture is true

$$b(B \vee \sim C \mid H) = b(C \vee \sim B \mid H) + b(B \mid H) - b(C \mid H), \qquad (23)$$

where *H* in this equation represents the issue to be resolved. Not surprisingly, this is a true equation and is easily proved using the algebra of questions or visualized in the Venn diagram of Figure 1b. While this logical notation may obscure the relation of this equation to Bayes' Theorem for assertions, it is much more easily interpreted in application [5].

## Symmetries

The symmetry between the logic of assertions and the logic of questions seems to possess more secrets. These relationships can be made clearer by looking again at the commutativity of the conjunction of assertions

$$p(b \wedge c \mid h) = p(b \mid h)\, p(c \mid h \wedge b) \qquad (24)$$
$$p(c \wedge b \mid h) = p(c \mid h)\, p(b \mid h \wedge c) \qquad (25)$$

which gives Bayes' theorem

$$p(b \mid h \wedge c) = p(b \mid h) \frac{p(c \mid h \wedge b)}{p(c \mid h)}. \qquad (26)$$

However, the sum rule could have been applied to (24) to obtain

$$p(b \wedge c \mid h) = p(b \mid h)\left(1 - p(\sim c \mid b \wedge h)\right) \qquad (27)$$
$$= p(b \mid h) - p(b \mid h)\, p(\sim c \mid b \wedge h) \qquad (28)$$
$$= p(b \mid h) - p(b \wedge \sim c \mid h). \qquad (29)$$

Applying the same procedure to (25) and equating the right-hand sides we get

$$p(b \wedge \sim c \mid h) = p(\sim b \wedge c \mid h) + p(b \mid h) - p(c \mid h), \qquad (30)$$

which is an alternate expression resulting from the commutativity of the conjunction of assertions.

The same game can be played with the commutativity of the disjunction of questions. As described above, we can derive Bayes' Theorem for questions

$$b(B \vee C \mid H) = b(B \mid H)\, b(C \mid B \vee H), \tag{31}$$
$$b(C \vee B \mid H) = b(C \mid H)\, b(B \mid C \vee H), \tag{32}$$

giving

$$b(B \mid C \vee H) = b(B \mid H) \frac{b(C \mid B \vee H)}{b(C \mid H)}, \tag{33}$$

which is analogous to Bayes' Theorem for assertions (26) above. Applying the sum rule for questions to (31) and (32) above we find

$$b(B \vee C \mid H) = b(B \vee \sim C \mid H) + b(C \mid H), \tag{34}$$
$$b(C \vee B \mid H) = b(\sim B \vee C \mid H) + b(B \mid H), \tag{35}$$

which gives

$$b(B \vee \sim C \mid H) = b(\sim B \vee C \mid H) + b(B \mid H) - b(C \mid H), \tag{36}$$

analogous to (30) above.

Notice that (36) is the equation that was previously suggested (via the relevance-entropy conjecture) by its information-theoretic counterpart (21) derived from Bayes' Theorem for assertions (26). We can in fact perform the same operations to obtain another interesting relation. First we take the logarithm of (31)

$$\log b(B \vee C \mid H) = \log b(B \mid H) + \log b(C \mid B \vee H), \tag{37}$$

followed by the expected value over all possible questions $B \vee C$ to obtain

$$\sum_{B,C} b(B \wedge C \mid H) \log b(B \wedge C \mid H)$$
$$= \sum_{B,C} b(B \wedge C \mid H) \log b(B \mid H) + \sum_{B,C} b(B \wedge C \mid H) \log b(C \mid B \wedge H) \tag{38}$$

By the sum rule for relevance, the sum over $C$ in the first term on the right-hand side marginalizes to one leaving only the sum over $B$. We define new functions $G$ such that the term of the left is $-G(B,C)$, the first term on the right is $-G(B)$ and the final term is $-G(C|B)$. Performing the same operations on (35) and equating the right-hand sides we obtain

$$G(B \mid C) = G(C \mid B) + G(B) - G(C). \tag{39}$$

As expected, this is similar in form to (30) above.

This leads us to conjecture that the probability of an assertion can be written in terms of the relevance of the questions that have that assertion as an answer. This can be written explicitly as

$$p(a \mid h) = G(\{A_i\}, H), \tag{40}$$

where $\{A_i\}$ is the set of all questions which have assertion $a$ as their answer. This is a novel conjecture analogous to the relevance-entropy conjecture, which can be written similarly

$$b(A|H) = H(\{a_i\},h), \qquad (41)$$

where $\{a_i\}$ is the set of all assertions answering question $A$. Note that the question $H$ in (40) should not be confused with the function $H(\cdot)$. In addition, $H(\cdot)$ and $G(\cdot)$ must have the same form with the usual assertion-question and conjunction-disjunction interchange. More importantly, these functions are not inverses of one another as they take a set of elements as arguments. Thus schematically we have the situation shown in Figure 2.

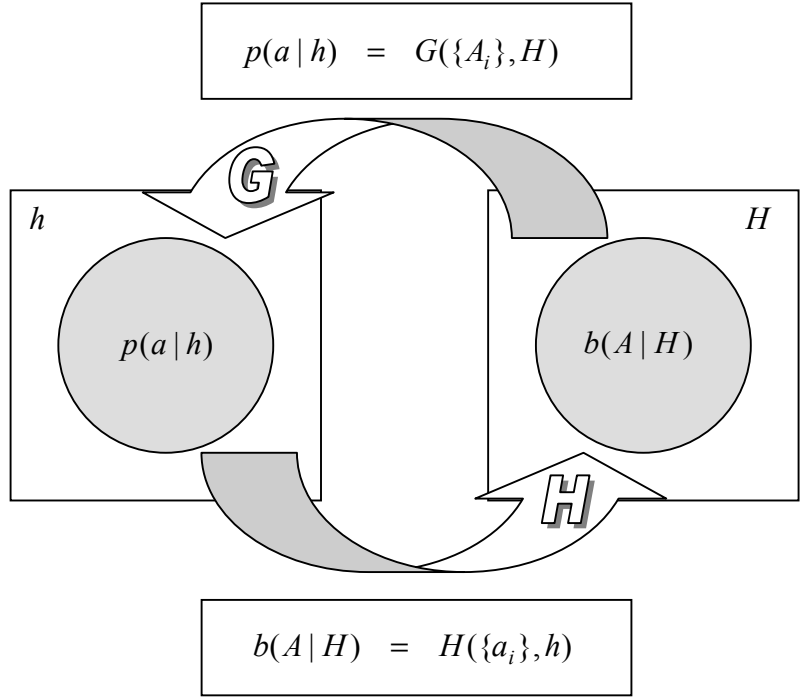

**FIGURE 2.** This is a cartoon depicting the conjectured relationship between the probabilities of assertions and the relevance of questions.

Finally, there are two forms of relations derived from commutativity in each of the two spaces: the product-form for assertions

$$p(b|h \wedge c) = p(b|h)\frac{p(c|h \wedge b)}{p(c|h)}, \qquad (26)$$

known as Bayes' Theorem, its associated sum-form for assertions

$$p(b \wedge \sim c|h) = p(\sim b \wedge c|h) + p(b|h) - p(c|h), \qquad (30)$$

the product-form for questions

$$b(B|C \vee H) = b(B|H)\frac{b(C|B \vee H)}{b(C|H)}, \qquad (33)$$

and its associated sum-form for questions

$$b(B \vee \sim C \mid H) = b(\sim B \vee C \mid H) + b(B \mid H) - b(C \mid H). \quad (36)$$

The product-forms in each space are analogous to one another, as are the sum-forms (with interchange of probability-relevance, assertion-question and conjunction-disjunction). Now the function $H$ takes the product-form for assertions to the sum-form for questions, while the function $G$ takes the product-form for questions to the sum-form for assertions.

## PRACTICALITIES

While some may find these theoretical issues interesting, others may be wondering if this viewpoint has any practical use. The multitudes of papers presented at previous Maximum Entropy and Bayesian Methods workshops have well demonstrated the power of inductive inference when applied to data analysis. Even more so, these previous works demonstrate the merit and utility of the general viewpoints of Jaynes and Cox regarding probability as representing the relative degree of implication among logical assertions. For this reason, I refer the interested reader to another source [6] for a detailed description of the process of data analysis using Bayesian or inductive inference.

There presently exist few applications that demonstrate inductive inquiry or inductive logic in general. Most notable are the works of Robert Fry [7-11]. In addition, this author has demonstrated the side-by-side application of inductive inference and inductive inquiry with application to a source separation problem [5].

### Experimental Design

The problem of experimental design has received much less attention than the problem of data analysis. This is perhaps because the logic of questions is much less understood than the logic of assertions. In terms of inductive inquiry, the problem statement and the form of its solution are straightforward. There exists an unresolved scientific issue of interest, $S$. For practical reasons this question cannot be asked directly, and we are forced to resort to asking an experimental question, $E$, in an attempt to resolve the issue. Out of all that can be asked, $H$, we focus our inquiry on the scientific issue, $S \vee H$. More relevant experimental questions are those that have greater relevance toward (or bearing on) this focused issue, written as $b(E \mid S \vee H)$. Using the product rule, we can write this relevance as

$$b(E \mid S \vee H) = \frac{b(E \vee S \mid H)}{b(S \mid H)}. \quad (42)$$

Finding the experimental question with maximal relevance requires maximizing this quantity with respect to all possible experimental questions. Note that the term in the denominator does not vary as different experimental questions are considered. Thus the most relevant experiment can be determined by maximizing the relevance of the common question $E \vee S$. If the relevance-entropy conjecture is in fact true, this is identical to maximizing the mutual information between the experimental question and the scientific issue. The process of experimental design could then be viewed information-theoretically as the process of designing a communication channel between the system of interest and the experimenter.

From (42), to maximize the relevance of the experimental question to the scientific issue, one must maximize the relevance of the common question

$$b(E \vee S \mid H) = b(S \mid H) + b(E \mid H) - b(E \wedge S \mid H). \tag{43}$$

By re-writing the conjunction on the right-hand side, we get

$$b(E \vee S \mid H) = b(S \mid H) + b(E \mid H) - \left(b(S \mid H) + b(E \wedge \sim S \mid H)\right), \tag{44}$$

which simplifies to

$$b(E \vee S \mid H) = b(E \mid H) - b(E \wedge \sim S \mid H). \tag{45}$$

This can be written in terms of the probabilities of the assertions that answer the experimental question and scientific issue. The possible answers to the experimental question are a set of statements describing the data that could be recorded. Any particular set of data will be denoted by the joint assertion $e_i$. Similarly, the possible answers to the scientific issue are a set of statements describing the possible models for the physical situation under consideration. Any particular model will be denoted by the joint assertion $s_j$.

If the relevance-entropy conjecture is correct, the relevances in (45) can be written in terms of the entropy of the probabilities of the experimental and scientific answers. This is equivalent to (in information-theoretical notation)

$$I(E;S) = H(E) - H(E \mid S), \tag{46}$$

which after simplification gives

$$b(E \vee S \mid H) = -\sum_i p(e_i \mid h) \log p(e_i \mid h) + \\ - -\sum_j p(s_j \mid h) \sum_i p(e_i \mid s_j \wedge h) \log p(e_i \mid s_j \wedge h). \tag{47}$$

Examining the terms on the right, one can see that this result is quite intuitive. To find an experimental question that has greatest relevance to the scientific issue at hand, one must choose an experiment that has two qualities. First, the experiment should maximize the entropy of the set of possible results (first term). In other words, the experiment should be maximally unbiased. Second, the entropy of the likelihood function summed over all possible scientific scenarios should be minimized (second term). This means that a good experiment will result in data that provides the sharpest estimates of the model parameters on average. While these ideas are not new to the problem of experimental design, seeing them derived here using inductive logic and the relevance-entropy conjecture is quite satisfying. For example, one standard technique in experimental design is to simulate one of the possible physical situations and choose an experimental design that minimizes the variance of the likelihood function [6]. This is an approximation to the second term derived above.

# OPEN QUESTIONS

Richard Cox's work was essential in putting modern probability theory on a firm ground based on sound logical principles. In addition, we were fortunate to have as his last work a glimpse into the logical duality between assertions and questions thus opening a broader scope: logical induction. This glimpse suggests a richer relationship between probability and entropy.

One of the current difficulties is that it is yet unclear how to completely relate these two spaces to one another. It is expected that the relevance, or bearing, of a question must be expressible in terms of the probabilities of the assertions that answer that question via some function $H$. While much evidence suggests that this function may be the entropy, this is not yet proven. However, something in this resonates with intuition. Probability describes the degree of certainty, whereas entropy describes the degree of uncertainty. Again we have what is known versus what is unknown. More unusual is the hypothesized relationship, denoted by the function $G$, which takes relevance to probability. The duality between the spaces suggests that this function $G$ has the same form as $H$, with the usual probability interchanged with relevance, assertions interchanged with questions and conjunctions interchanged with disjunctions. However, this actually depends on whether the space of assertions is isomorphic to the space of questions. This is not immediately obvious even given the duality we have explored. Clearly more investigation is needed to fully explore the structure of these spaces.

Finally, playing with questions proves to be quite difficult at first, as the intuition seems to be lacking.[5] This may explain our reliance on the function $H$, which translates everything back to assertions where we are more comfortable. However, it should be possible to derive quantities like prior relevance and to perform calculations in question space without ever resorting to assertions. There may be some fascinating research here.

# ACKNOWLEDGMENTS

I would like to thank Robert Fry for introducing me to Richard Cox's work in inductive inquiry and for exciting discussions on potential applications of this theory.

---

[5] Guillaume Marrelec and I spent several hours during the workshop trying to get our heads around a simple problem dealing with the relations between the questions one can ask regarding the state of a four-sided die.